# Machine Learning Prediction of Charged Defect Formation Energies from Crystal Structures

Shin Kiyohara,[1] Chisa Shibui,[1] Soungmin Bae[1] and Yu Kumagai[1,2,*]

[1] *Institute for Materials Research, Tohoku University, Sendai, Japan*

[2] *Organization for Advanced Studies, Tohoku University, Sendai, Japan*

*Corresponding author: yukumagai@tohoku.ac.jp

**Recent advances in materials informatics have expanded the number of synthesizable materials. However, screening promising candidates, such as semiconductors, based on defect properties remains challenging. This is primarily due to the lack of a general framework for predicting defect formation energies in multiple charge states from structural data. In this Letter, we present a protocol, namely data normalization, Fermi level alignment, and treatment of perturbed host states, and validate it by accurately predicting oxygen vacancy formation energies in three charge states using a single model. We also introduce a joint machine-learning model that integrates defect formation energies and band-edge predictions for virtual screening. Using this framework, we identify 89 hole-dopable oxides, including BaGaSbO, a potential ambipolar photovoltaic material. Our protocol is expected to become a standard approach for machine-learning studies on point defect formation energies.**

Historically, superior materials that have contributed to human society's development have been discovered through experimental efforts. However, the cost and time required for such experiments pose significant constraints on material innovation. To mitigate these challenges, first-principles calculations have been adopted in materials science over the past few decades, especially thanks to the exponential growth in computational power [1]. Nowadays, high-throughput computational techniques have facilitated the development of extensive materials databases. They are currently employed in building machine learning (ML) surrogate models, which are then used to screen promising materials from millions of candidates [2].

Despite these advances, the number of theoretically identified materials that achieve commercial viability remains limited. One fundamental reason is that theoretical calculations usually assume defect-free materials, which can diverge from practical conditions where defects often play a crucial role, particularly in semiconductors. For instance, desirable photoabsorbers for solar cells should be free from point defects with deep levels [3]. In addition, the dopability may be limited by the compensation of native defects [4]. Thus, integrating defect-property-based screening into computational materials discovery is expected to improve the chances of identifying commercially viable semiconductors.

To this end, Deml *et al*. constructed a linear regression model for the formation energies of neutral oxygen vacancies ($V_O$) in 45 binary and ternary oxides [5]. We also systematically calculated $V_O$ not only in neutral charge state but in +1, and +2 charge states in ~1000 oxides that are composed of diverse crystal structures and cationic elements [6] and applied random forest (RF) regression to predict the $V_O$ formation energies, hereafter refer to as $E_f[V_O]$. Subsequently, Witman et al. also calculated the neutral $V_O$ in approximately 200 oxides and performed ML using graph-based neural network (GNN) models to predict their formation energies [7]. Frey et al. also calculated neutral defects in 158 two-dimensional materials and constructed a GNN model that predicts formation energies of defects [8] . While other research groups have also developed ML models for predicting defect formation energies, the atomic frameworks were fixed to specific types, such as perovskite [9] and zinc-blende structures [10,11].

Virtual screening requires ML models that predict defect formation energies in *various charge states* using *structural information alone*. In addition, when screening superior materials from a myriad of candidates, the models must accommodate *diverse crystal structure*s. Furthermore, a single ML model should be able to predict multiple defect-charge combinations, analogous to a general-purpose ML potential. However, prior regression models fall short in these criteria. For example, our RF models require descriptors obtained from first-principles calculations for unit cells, such as dielectric constants and O-2*p* center positions [6]. Also, the models developed by several groups focuses only on the neutral defects [5,7,8].

Indeed, a general ML framework for defect formation energies at various $q$ values has not yet been established. Our purpose is to present one here using $E_f[V_O]$ as a case study. The key aspects include pruning inadequate data, normalization of the defect formation energies with different $q$ values, and determination of the Fermi level ($\epsilon_F$), which linearly shifts formation energies of charged defects. In addition, we discuss how to manage the defects with perturbed host states (PHS). We confirmed that all these factors improve the prediction of $E_f[V_O]$ by crystal graph convolution neural network models (CGCNN) [12] Using a single ML model and only structural information, we achieved improved accuracies of 0.29, 0.22, and 0.37 eV for the neutral, +1, and +2 charge states, respectively, outperforming our previous models. We further introduce an ML model that predicts band edge position and apply a joint model combining defect and valence-band maximum (VBM) ML models to virtually screen particularly rare hole-dopable oxides among recently proposed stable structures [13–15]. Consequently, we identified 89 hole-dopable oxides, including BaGaSbO as a potential ambipolar photovoltaic material.

*Charged defect formation energies.*

Since defect formation energies depend on $\epsilon_F$ when $q \neq 0$, we must align $\epsilon_F$ across compounds. The simplest choice is to set $\epsilon_F = \epsilon_{\mathrm{VBM}}$ of each compound. However, this is inadequate because $\epsilon_{\mathrm{VBM}}$ varies significantly across materials. For example, oxides with valence bands composed of lone-pair orbitals or transition-metal *d* orbitals generally have higher VBMs than those dominated by O-2*p* orbitals [6]. Therefore, defect formation energies at $\epsilon_F = \epsilon_{\mathrm{VBM}}$ reflect both information on the defect formation energy and $\epsilon_{\mathrm{VBM}}$, complicating prediction using a *single* GNN. Instead, we propose using core potentials as references. When a homogeneous charge is placed in the core potential region, the electrostatic energy remains constant under this alignment. Thus, aligning $\epsilon_F$ using core potentials ensures consistent electrostatic contributions to the defect formation energies (Fig. S1 in Supplemental Information, SI).

However, arbitrariness remains in determining $\epsilon_F$. During training, standardizing target values is known to improve weight updates [16]. A difficulty arises when training on formation energies of defects with mixed charge states. Figure 1(a) shows the distributions of $E_f[V_{\mathrm{O}}]$ at $q$=0, +1, and +2, where $\epsilon_{\mathrm{F}}$ of all the oxides are set to the VBM of ZnO via oxygen core potential alignment; ZnO was selected as the reference example, while noting that the choice of the Fermi level is, in principle, arbitrary. The energy distributions for different $q$ depend on $\epsilon_{\mathrm{F}}$; for example, as $\epsilon_{\mathrm{F}}$ decreases from the VBM of ZnO, each energy distribution becomes more separated. To train the GNN effectively, the distributions should overlap as much as possible. We define the difference in means ($\Delta$) across charge states as:

$$\Delta(\epsilon_F)=\sum_q \sum_{q'>q} \left|\mu(q,\, \epsilon_F)\text{-}\mu(q',\, \epsilon_F)\right|, \tag{1}$$

where $\mu$ is the mean of the dataset at given $q$ and $\epsilon_F$. For $V_{\mathrm{O}}$, $q$ and $q$' can be 0, +1 or +2. We choose an arbitrary $\epsilon_F$ to minimize $\Delta$, then constantly shift all $E_f[V_{\mathrm{O}}]$ and standardize each distribution. As shown in Fig. 1(b), the resulting distributions overlap well, as desired.

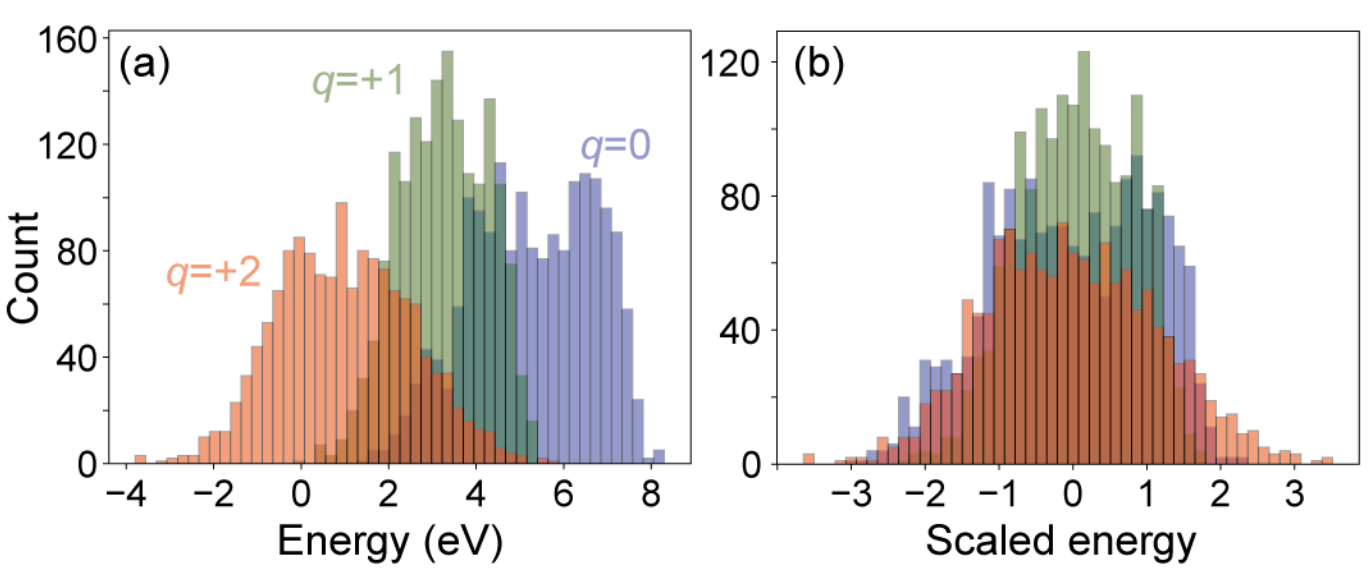

Fig. 1 (a) Distributions of oxygen vacancy formation energies at $q$=0, +1, and +2 with $\epsilon_F$ aligned to the VBM of ZnO via core potentials. Data for the vacancies with perturbed host states are already removed. (b) Normalized distributions with minimized mean differences between charge states (see text for details).

*Perturbed host states*. We here focus on perturbed host states (PHS) using donor-type defects as examples, which has rarely been discussed despite its significance; the same applies to acceptor-type defects. Figure 2 schematically illustrates eigenvalues in supercell models with defects. When defects exhibit localized in-gap states, they remain spatially confined and can be described using supercells composed of ~100 atoms (Fig. 2(a)). However, when defects possess the localized occupied states above the conduction band minimum (CBM) (Fig. 2(b)), the electrons drop into the CBM and become loosely trapped by defect centers electrostatically, which are called PHS or shallow states [17–19] . Their weak binding causes them to extend over thousands of atoms or more [20] .

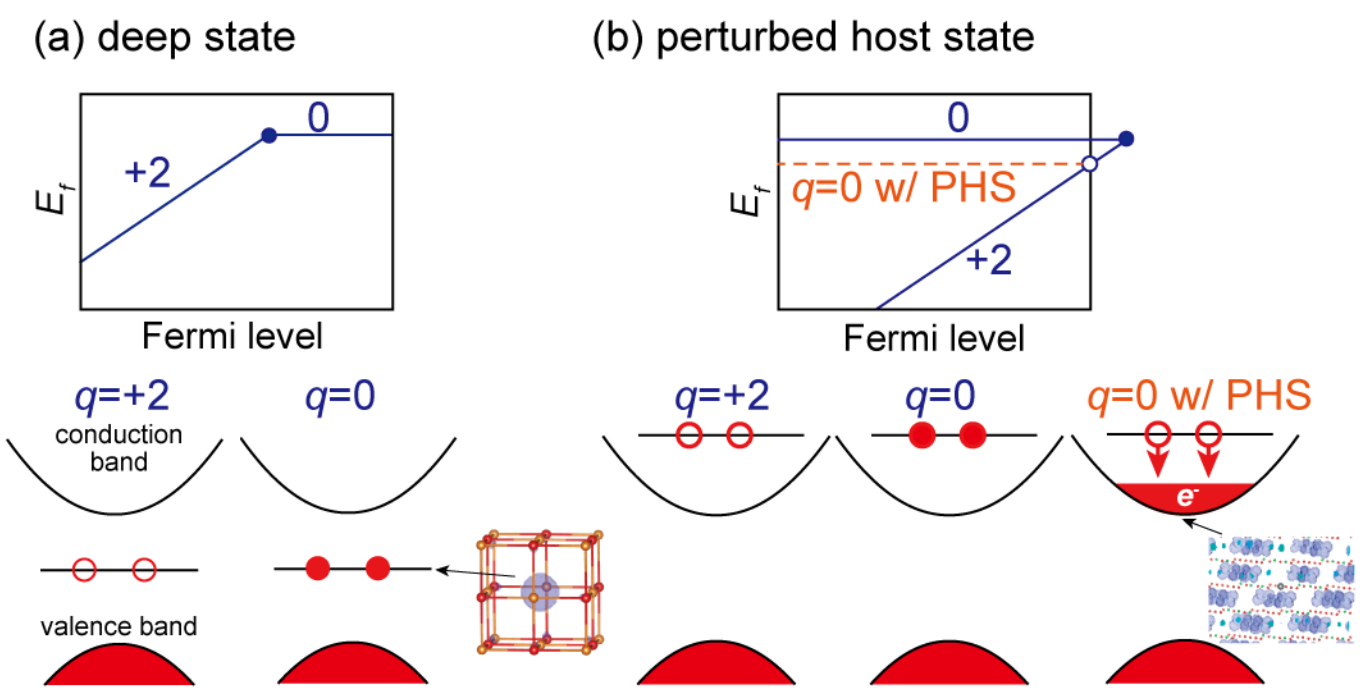

Fig. 2 Schematic illustration of defects with localized occupied states (a) inside the band gap and (b) above the conduction band minimum (CBM). In (b), the electrons are consequently dropped to the CBM, resulting in the delocalized perturbed host states (PHS). The formation energy of a defect with PHS is indicated by a dashed line. Donor transition levels are schematically indicated in the formation energy diagrams as a function of the Fermi level. Transition levels involving deep localized states are accurately calculated from first principles (solid circles). In contrast, shallow levels related to the PHS are shown qualitatively (open circle), as they cannot be computed with realistically sized supercells (see text for details). Squared wave functions of these states in neutral oxygen vacancies in MgO and $BaTiO_3$ are also shown as examples.

The formation energy of a defect $D$ in charge state $q$ ($D^q$) with a single PHS, calculated using a supercell, is given by:

$$E_f[D^q]|_{\epsilon_F=\epsilon_{\mathrm{CBM}}} = E_f\left[D^{q+1}\right]|_{\epsilon_F=\epsilon_{\mathrm{CBM}}} + E_{\mathrm{bind}} + E_{\mathrm{overlap}} + \int_{\epsilon_{\mathrm{CBM}}}^{\infty} (\epsilon - \epsilon_{\mathrm{CBM}}) D(\epsilon) f(\epsilon)\, \mathrm{d}\epsilon + \epsilon_{\mathrm{CBM}}, \quad (2)$$

where $\epsilon_{\mathrm{CBM}}$ is the eigenvalue of the CBM, and $E_f[D^q]|_{\epsilon_F=\epsilon_{\mathrm{CBM}}}$ is the formation energy of $D^q$ at $\epsilon_F = \epsilon_{\mathrm{CBM}}$. $D$ and $f$ are the density of states and orbital occupation fraction in the *supercell*, respectively. Note that the first term on the right-hand side is a defect formation energy without PHS at $\epsilon_F = \epsilon_{\mathrm{CBM}}$. The second term corresponds to the binding energy of the carrier electron to the positively charged defect center, typically ranging from a few tens to a few hundred meV [20] . The third term represents the energy change from the overlap between PHSs in a supercell with periodic boundary conditions, generally negative, analogous to the stabilization observed in alkali metals composed of positively charged ion cores and free electrons. The fourth term arises from Burstein-Moss-like (BM) effects when multiple $k$-point sampling is used, being positive due to electron occupancy above the CBM. Note that the third and fourth terms arise from using small supercells, making accurate calculation of these small binding energies infeasible. In some studies, the formation energies of defects with PHS have been evaluated by removing the fourth term [21–23]; however, this is insufficient due to the contribution of the third term, which can correspond to a few hundred meV, resulting in the deep transition levels of the PHS [17–19] . Therefore, we usually present PHS-related transition levels qualitatively [14,24,25] , as shown by an open circle in Fig. 2(b). Note that when the formation energies of deep-level defects are accurately predicted by ML models, the PHS-related transition levels can still be qualitatively described, as the transition levels are predicted to lie above the CBM, as shown in Fig. 2(b).

To simplify the discussion, we ignore the second and third terms, which are on the order of a few tenths of an eV and assume that the PHS occupies the CBM (*i.e.*, no band filling effect). Then, we obtain $E_f[D^q]|_{\epsilon_F=\epsilon_{\mathrm{CBM}}}=E_f\left[D^{q+1}\right]|_{\epsilon_F=\epsilon_{\mathrm{CBM}}}+\epsilon_{\mathrm{CBM}}$ . This shows that $E_f[D^q]$ consists of $E_f\left[D^{q+1}\right]$ and the single-particle level of the CBM, meaning that the defect formation energies with PHS qualitatively differ from those without PHS. Indeed, excluding defects with PHS from the dataset improves prediction accuracy, as shown later.

*Accuracy evaluation*. To validate our data handling framework, we built a GNN model using an $E_f[V_{\mathrm{O}}]$ database with mixed $q$ values [6] . Crystal structures are modeled using the CGCNN framework, as shown in Fig. S2 in SI (see also *Method* for details). Other GNN models can also be used; the choice usually depends on the dataset size. We firstly encoded the atomic species and bond lengths, then passed them to convolutional layers. Subsequently, we extracted features at O sites at a pooling layer, and concatenated the charge state $q$. Finally, the data were fed into a fully connected neural network to predict an $E_f[V_{\mathrm{O}}]$. (see Methods and Fig. S2 in SI).

Figure 3(a) shows the parity plot of $E_f[V_{\mathrm{O}}]$ for the test sets with the average mean absolute errors (MAEs). The prediction errors are 0.29, 0.22 and 0.37 eV for $q$=0, +1, and +2, respectively, which are lower than in our previous study [6], despite using only structural information and a single GNN model here. Note that learning defect formation energies for each charge state $q$ may improve accuracy, but reduces generality, as ML models must be built for every element–charge state combination. We confirmed that including data with PHS reduces accuracy by 0.02–0.03 eV (Fig. S3 in SI), even though the dataset size increases by 10%.

The formation energies of charged defects depend linearly on the Fermi level, which varies within the band gap. Therefore, the band edge positions obtained from core potentials in unit cell calculations must also be determined. Since unit cells are not computed in structure-based virtual screening, separate ML models to predict the band edges are needed. In this study, we constructed a CGCNN model on a VBM dataset for the same oxides used in constructing oxygen vacancy database. Figure 3(b) shows a parity plot for the VBM, showing high accuracy of our CGCNN model.

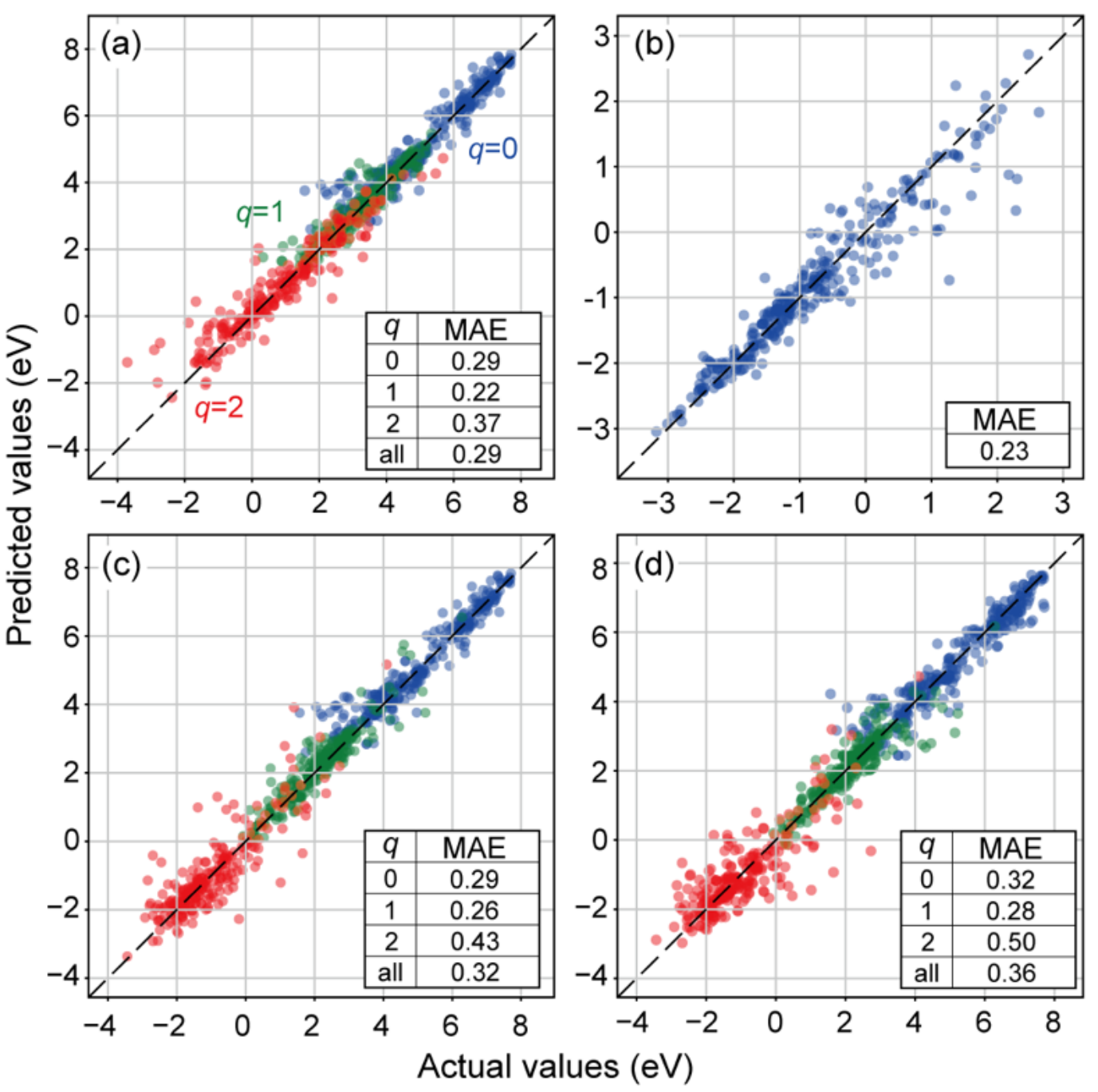

Fig. 3 Parity plots of (a, c, d) oxygen vacancy formation energies ($E_f[V_\mathrm{O}]$) and (b) VBM positions. (a) $E_f[V_\mathrm{O}]$ with the Fermi level ($\epsilon_\mathrm{F}$) aligned to the VBM of ZnO. (c, d) $E_f[V_\mathrm{O}]$ with $\epsilon_\mathrm{F}$ aligned to the VBM of each compound, predicted using (c) a joint model and (d) a single model (see text for details). The $x$-axis shows values from first-principles calculations, while the $y$-axis shows our crystal graph convolution neural network predictions. Results are for test sets composed of 140 oxides. Mean absolute errors are shown in eV in the insets. Due to site-dependent core potentials, the VBM in (b) also vary by site.

Using the CGCNN models for both $E_f[V_\mathrm{O}]$ and VBM, referred to as a joint model, we predicted $E_f[V_\mathrm{O}]$ at the VBMs, as shown in Fig. 3(c). For comparison, Fig. 3(d) shows $E_f[V_\mathrm{O}]$ predicted directly by a single CGCNN model based on the datasets of $E_f[V_\mathrm{O}]$ at the VBM. The joint model achieves higher accuracy, with slight improvements of 0.02–0.07 eV. It is, however, noted that since the VBM can be obtained from unit cell calculations, unlike point defects, its prediction accuracy can be significantly improved using large-scale datasets like the Materials Project [2] ,

which is a major advantage of this joint model.

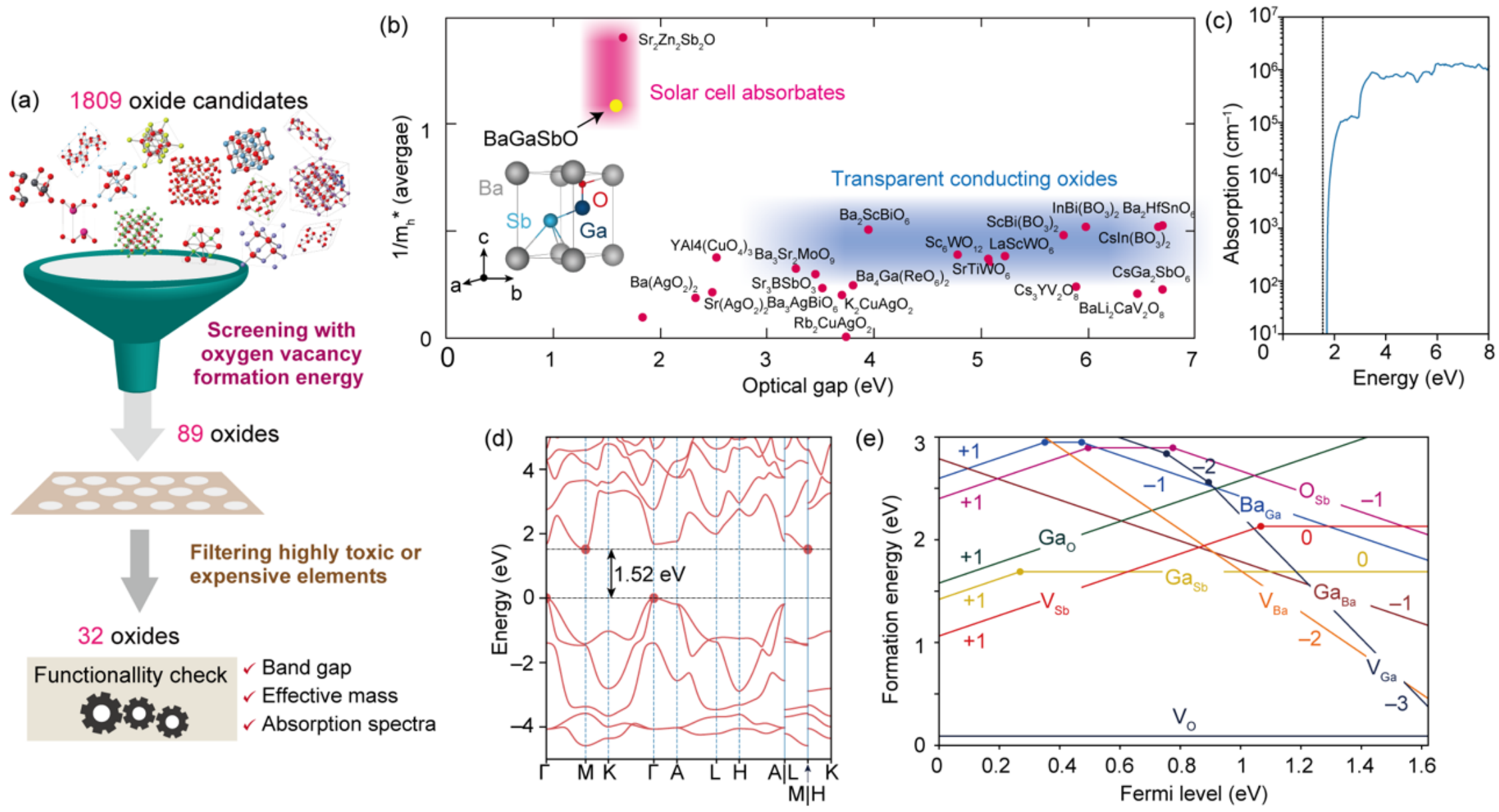

Fig. 4 (a) Schematic illustration of our screening process for *p*-type oxides. (b) Reciprocal of the averaged hole effective mass ($m_h$*) in units of the free-electron mass plotted against the optical gap for 32 screened oxides. Candidates for solar cell absorbers and transparent conducting oxides are highlighted. The crystal structure of BaGaSbO is also shown in the inset. (c) Optical absorption coefficient and (d) band structure of BaGaSbO. The band edge positions are marked with filled circles. (e) Formation energies of native defects in BaGaSbO under the Ga-rich condition. The charge states are also described. $V_i$ denotes a vacancy of element *i*. These calculations were performed using the dielectric-dependence hybrid functionals (see the main text for details).

*Virtual screening.* Using the joint model, we screened as-yet-unsynthesized hole-dopable oxides, since hole doping is typically hindered by compensation from positively charged $V_O$ [14] . Our targets are the stable oxides predicted by Merchant et al. [26] . We firstly screen the oxides with the same criteria as the training set, identifying 1,809 candidate oxides (Fig. 4(a)). Using our joint model, we screened oxides where $V_O$ do not act as hole killers, that is, all $V_O$ are stable in neutral or positive charge states with positive formation energies at the VBM in the O-rich conditions. The number of remaining oxides is 89 (4.9%), indicating how difficult to make *p*-type oxides.

To identify promising *p*-type oxides for applications, we excluded oxides containing highly toxic or expensive elements, leaving 32 candidates (see *Methods*). For these, we calculated the electron and hole effective masses, electronic and ionic dielectric constants, and optical absorption spectra using dielectric-dependent hybrid (DDH) functionals [27,28], which accurately predict the band gaps. Consequently, we have identified several prospective oxides as solar cell absorbates and transparent conducting oxides as shown in Fig. 4(b).

BaGaSbO is a promising photovoltaic material among them. As shown in the inset of Fig. 4(b), it exhibits very low electron and hole effective masses in the *ab*-plane (0.34 and 0.19$m_0$, where $m_0$ is the free-electron rest mass, respectively), a steep optical absorption onset (Fig. 4(c)) and slight indirect gap (Fig. 4(d)), which mitigates carrier recombination. Although $Sr_2Zn_2Sb_2O$ exhibits more favorable effective masses, its large indirect-direct band gap difference (0.7 eV) would unfortunately cause significant efficiency loss (see Fig. S4 in SI).

Figure 4(e) presents the calculated defect formation energies in BaGaSbO with the DDH functional. As predicted by our join model, the oxygen vacancies are stable in the neutral charge state even at $\epsilon_F = \epsilon_{\mathrm{VBM}}$, indicating no hole compensation. Furthermore, there are no defects with both low formation energies and deep levels, and it also show *p*-type (*n*-type) behavior using Mg or Zn (La or Y) as dopants, respectively (Fig. S5 in SI), indicating its remarkable potential as a photovoltaic material.

*Conclusion*. This study introduces an ML framework for point defect formation energies across multiple charge states. Focusing on $V_{\mathrm{O}}$, we demonstrate how to prune inadequate data and normalize formation energies using the $\epsilon_F$ alignment. Our single CGCNN model achieves high accuracy merely with structural information. We further introduce the joint model that combine the models predicting the formation energies of defects and VBMs. With this model, we virtually screened stable oxides and identified 89 hole-dopable candidates, highlighting BaGaSbO as a promising ambipolar semiconductor for solar energy applications.

Finally, we discuss the inherent limitations and the scope of applicability of the proposed procedure. In the present protocol, the Fermi level is aligned using the core potentials. This procedure is also applicable to antisite defects and extrinsic substitutional dopants, since the core potentials of the original atoms in the unit cell are well-defined. In contrast, for interstitial defects, no atoms occupy the corresponding positions in the unit cell, and thus the protocol cannot be directly applied in such cases. In addition, the error introduced by the ML model adds to the intrinsic errors of first-principles calculations. Therefore, higher-level reference data, such as those obtained from hybrid-functional calculations, are desirable to reduce the overall inaccuracy.

The ML model is not intended to replace first-principles calculations but rather to complement them, serving as a prescreening tool to identify promising materials among numerous candidates in which point defects play crucial roles, as demonstrated in our screening of hole-dopable oxides.

*Methods*. To test the performance of our protocol, we used an $V_O$ database comprising non-magnetic 932 oxides with 2090 inequivalent sites [6]. Note that the PBEsol functional [29] with Hubbard $U$ corrections [30] for Cu and Zn $d$ orbitals and Ce $f$ orbitals with $U_{eff} = 5$ eV was used to estimate the formation energies of oxygen vacancies, while the band edge positions were determined using the non-self-consistent DDH functional [27]. Note that, although the spin–orbit coupling was not taken into account, our CGCNN model can be combined with the band-edge positions determined with spin–orbit coupling in an *ad hoc* manner.

Initially, we excluded $V_O$ that migrate to different sites or accompany neighboring atoms to higher-symmetry interstitials. We also removed $V_O$ in dynamically unstable host structures (see Supplemental Note 1 in SI). The numbers of $V_O$ with $q$ = 0, +1, and +2 are 1734 (2045), 1676 (2014), and 1416 (1637) after (before) removing defects with PHS. To identify $V_O$ with PHS, we used the algorithm based on eigenvalues and orbital components of the single-particle levels near the band edges [6]. To avoid the data leakage, we divided the dataset by the oxides rather than by the O sites as we did previously [6]. The dataset was split into training, validation, and test sets in a 0.7:0.15:0.15 ratio for evaluating the accuracies. When constructing the CGCNN models, elemental information is embedded using original CGCNN descriptors, and bonding is represented via Gaussian filters (see Supplemental Note 2 in SI) [7] . The model is illustrated in Fig. S2 and the tuned hyperparameters with optuna [31] are also tabulated in Tables S1 and S2 in SI. $E_f[V_O]$ was evaluated by setting the oxygen chemical potential to half the total energy of an $O_2$ molecule in the spin-triplet state.

For virtual screening, the model was retrained on the full dataset using the same hyperparameters. In the structural database provided by Merchant *et al.*, [26] slightly metastable structures are also included. Here, however, we focus on oxides stable against competing phases to enhance the likelihood of synthesizing the predicted candidates. For the screening by $E_f[V_O]$, we assumed oxygen-rich conditions. To determine these, we retrieved the total energies of the competing phases from Materials Project [2] and constructed the chemical potential diagrams accordingly. In the next step, compounds with highly toxic (Tl, Hg, As, Cd, Pb, Cr) or expensive (Au, Pt, Pd) elements are excluded. We used DDH to evaluate the band gaps of screened oxides and the defect formation energies in BaGaSbO. Details of the first-principles calculations are described in the Supplemental Note 3 in SI [28,32–37].

The data and code used in this study is available at [38].

*Acknowledgments.* This work has been supported by JST FOREST Program (JPMJFR235S), and KAKENHI (22H01755 and 25K01486).